\documentclass[twocolumn, showpacs, prl]{revtex4-1}

\usepackage{graphicx}

\usepackage{dcolumn}

\usepackage{bm}

\usepackage[usenames]{color}

\begin{document}

\title{Ferromagnetic fractional quantum Hall states in a valley degenerate
\newline{ two-dimensional electron system }}

\date{\today}

\author{Medini Padmanabhan}

\author{T.\ Gokmen}

\author{M.\ Shayegan}

\affiliation{Department of Electrical Engineering, Princeton
University, Princeton, NJ 08544}

\begin{abstract}

We study a two-dimensional electron system where the electrons
occupy two conduction band valleys with anisotropic Fermi contours
and strain-tunable occupation. We observe persistent quantum Hall
states at filling factors $\nu = 1/3$ and 5/3 even at zero strain
when the two valleys are degenerate. This is reminiscent of the
quantum Hall ferromagnet formed at $\nu = 1$ in the same system at
zero strain. In the absence of a theory for a system with
anisotropic valleys, we compare the energy gaps measured at $\nu =
1/3$ and 5/3 to the available theory developed for single-valley,
two-spin systems, and find that the gaps and their rates of rise
with strain are much smaller than predicted.

\end{abstract}


\maketitle

In a single-electron picture, there should be no quantum Hall effect
(QHE) at odd-integer Landau level (LL) filling factors ($\nu$) in a
two-dimensional electron system (2DES) with a vanishing Lande
$g$-factor. In an interacting 2DES, however, a QH state exists at
$\nu = 1$ even in the limit $g = 0$ \cite{sondhiPRB93,maudePRL96}.
The ground state is a ferromagnet, and it is the large Coulomb
(exchange) energy cost of a spin flip that leads to a sizable energy
gap separating this ground state from its excitations. These charged
excitations, called skyrmions, have a nontrivial, long-range spin
texture with a slow rotation of the spin as a function of distance
to minimize the Coulomb energy cost
\cite{sondhiPRB93,barrettPRL95,schmellerPRL95,maudePRL96,shuklaPRB00}.

A question naturally arises as to whether there are similar
ferromagnetic ground states in the $fractional$ quantum Hall effect
(FQHE) regime. The presence of such a state with finite energy gap
at $g$ = 0 along with skyrmionic excitations was theoretically
predicted for $\nu = 1/3$ \cite{rezayiPRB91,sondhiPRB93}.
Experimentally, it is challenging to tune the $g$-factor to zero
while keeping the quality of the 2DES sufficiently high to exhibit
FQHE. There has been only one study \cite{leadleyPRL97} to date
probing the state at $\nu = 1/3$ under the condition of $g
\rightarrow 0$, which was achieved via applying hydrostatic
pressure. The data revealed no FQH resistance minimum at $\nu = 1/3$
when $g \simeq 0$ \cite{footnote1}, but from measurements at
non-zero values of $g$, a finite but small energy gap was deduced
for the $\nu = 1/3$ FQH state.

\begin{figure}
\includegraphics[scale=1]{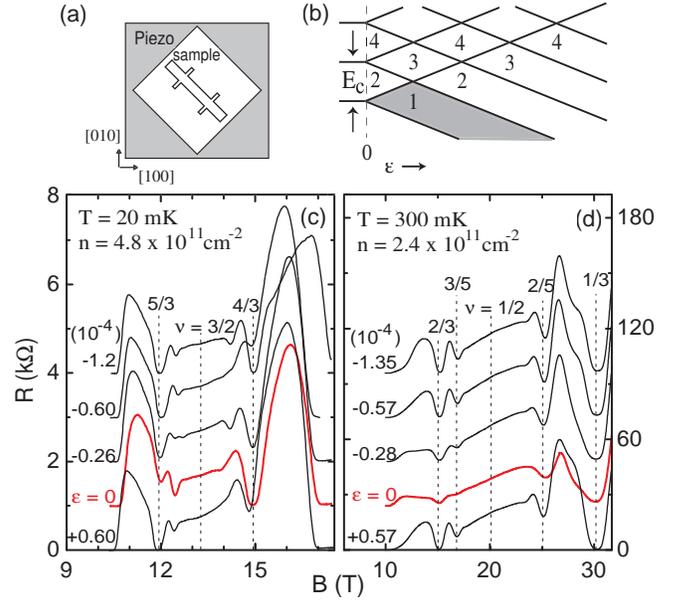}
\caption{(a) Experimental setup with the sample glued to a piezo.
(b) Simple energy level fan diagram showing the splitting of
electron LLs in response to strain. (c) and (d) Magnetoresistance
traces for sample A around $\nu = 3/2$ and $\nu = 1/2$ taken at
different $\epsilon$.}
\end{figure}

\begin{figure*}[t]
\centering
\includegraphics[scale=1]{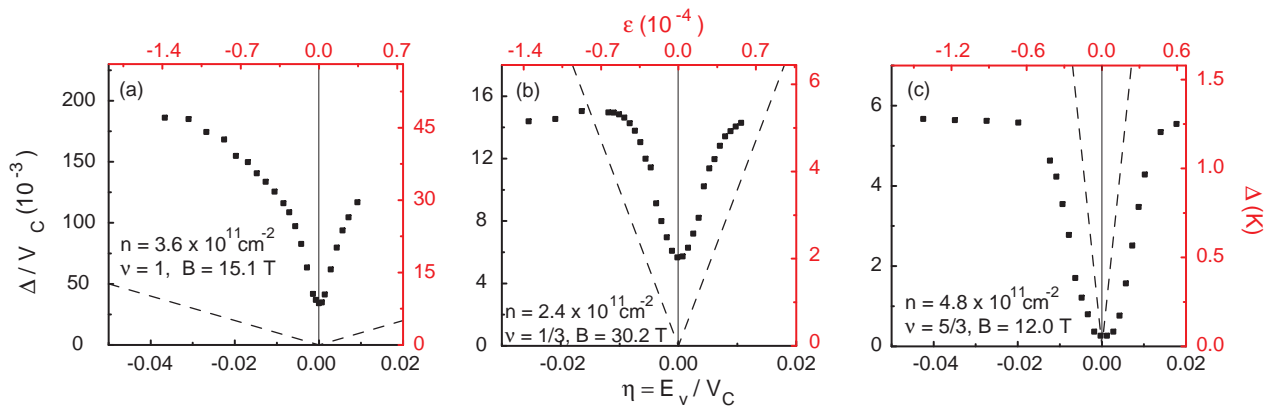}
\caption{Energy gaps vs. $\epsilon$ for sample A for (a) $\nu = 1$,
(b) $\nu = 1/3$, and (c) $\nu = 5/3$. Both the $x$ and $y$ axes are
also indicated normalized to the Coulomb energy, $V_{C}$. The
single-particle prediction, $\Delta = \epsilon E_{2}$ (with $E_{2}$
= 5.8 eV) is shown as dashed lines.}
\end{figure*}

Here we report magnetotransport measurements for 2DESs confined to
AlAs quantum wells where the electrons occupy two conduction band
minima (valleys). In this system, the energy splitting between the
two valleys can be controlled \textit{in situ} via applying in-plane
strain. Considering the valley degree of freedom as an isospin, the
condition of zero valley splitting is identical to the limit of $g =
0$ in a two-spin system. Under this condition, a ferromagnetic
ground state and skyrmionic excitations have already been reported
at $\nu = 1$ \cite{shkolnikovPRL05,shayeganPhysicaB06,footnote2}.
Here we report the observation of a well-developed QH resistance
minimum at $\nu = 1/3$ as we tune the valley splitting through zero.
We also observe a pronounced resistance minimum at $\nu = 5/3$ which
is the particle-hole conjugate state of $\nu = 1/3$ in our
two-component (two-valley) system. These observations strongly
suggest the presence of ferromagnetic FQH states at these fillings.
From the temperature ($T$) dependence of the resistance minima, we
deduce energy gaps for these states as a function of applied strain.
The gaps increase initially and then saturate, as expected
qualitatively. Quantitatively, however, both the gaps and their
dependencies on strain are much weaker than the predictions of
theories which were developed for the single-valley, two-spin
systems.

Our samples are AlAs quantum wells grown by molecular beam epitaxy
\cite{shayeganPhysicaB06}. Contacts were made with GeAuNi alloy and
Hall bars were defined. Metallic front and back gates were deposited
on the samples, thus allowing us to control the 2D electron density,
$n$. We report results for three samples from two wafers. The
quantum well in sample A is 12 nm wide while samples B and C have
well widths of 15 nm each. Measurements were done in either a 20 mK
base temperature dilution refrigerator or a 300 mK $^3$He system
using standard lock-in methods.

The 2D electrons in our samples occupy two conduction band valleys
with elliptical Fermi contours \cite{shayeganPhysicaB06}. In our
system, the Zeeman energy ($E_{Z}$) of electrons is typically larger
than their cyclotron energy \cite{shayeganPhysicaB06}. Since the FQH
states around $\nu = 1/2$ and 3/2 are formed in the first and second
LLs respectively, ours is a single-spin, two-valley system. The
valley degeneracy can be controllably lifted by uniaxial strain
applied in the plane of the sample. Experimentally, this is achieved
by gluing the sample to a piezoelectric (piezo) stack which expands
in one direction and contracts in the perpendicular direction when
an external voltage is applied. A schematic is shown in Fig.\ 1(a).
The single-particle energy splitting between the two in-plane
valleys is given by $E_{v} = \epsilon E_{2}$; $\epsilon =
\epsilon_{[100]}-\epsilon_{[010]}$ where $\epsilon_{[100]}$ and
$\epsilon_{[010]}$ are the strains along the [100] and [010] crystal
directions \cite{shayeganPhysicaB06} and $E_{2}$ is the deformation
potential, which has a band value of 5.8 eV in AlAs. This dependence
of the energy levels on strain is shown schematically in Fig. 1(b).

In Figs.\ 1(c) and 1(d) we show magnetoresistance traces taken for
sample A at different $\epsilon$. Figure 1(c) focuses on the
evolution of various FQH states around $\nu = 3/2$. In Fig.\ 1(d) we
present data from a different cooldown showing states around $\nu =
1/2$. The varying strengths of the different FQH resistance minima
around $\nu = 3/2$ as a function of $\epsilon$ have been explained
by the concept of composite fermions with a valley degree of freedom
\cite{bishopPRL07,padmanabhanPRB09}. Here, we focus on the states at
$\nu = 1/3$ and 5/3. The QH minima at both these fractions are
well-developed even when $\epsilon = 0$ indicating the presence of a
finite energy gap. This strongly suggests ferromagnetic ground
states at $\epsilon = 0$ at these fractional fillings.

In Figs. 2(b) and 2(c) we show a summary of our measured energy gaps
($\Delta$) for $\nu = 1/3$ and 5/3 as a function of strain-induced
valley splitting $E_v = \epsilon E_{2}$. These plots present the
highlights of our study. For comparison, we also include in Fig.
2(a) a plot of the measured gaps for the $\nu = 1$ QH state for the
same sample. In all three plots, the $x$ and $y$ axes are also shown
normalized to the Coulomb energy, $V_{C} = e^{2}/4 \pi \kappa
\epsilon_{0}\textit l_{B}$ where $\kappa = 10$ is the dielectric
constant of AlAs, $\textit l_{B}=\sqrt{\hbar/eB}$ is the magnetic
length and $B$ is the magnetic field. This normalization allows
easier comparison of our results with theoretical calculations. As
seen in Figs. 2(b) and 2(c), there are finite gaps at $\epsilon = 0$
at both $\nu = 1/3$ and 5/3. This is consistent with the
well-developed FQH minima observed in Fig. 1 and is the signature of
a QH ferromagnet.

Before discussing Fig.\ 2 in detail, we present data used to deduce
the gaps reported in Figs.\ 2(b) and 2(c). In Fig.\ 3 we show the
$T$-dependence of magnetoresistance traces taken around $\nu = 1/2$.
Figure 3(a) corresponds to the condition of $\epsilon = 0$ while
Fig. 3(b) is for $\epsilon = -1.35 \times 10^{-4}$ which belongs to
the saturated region in Fig. 2(b). The corresponding Arrhenius plots
used to deduce the gaps at $\nu = 1/3$ are shown as insets. Figure 4
shows data for the case of $\nu = 5/3$. Figure 4(a) contains
examples of Arrhenius plots at $\nu = 5/3$ for sample C for three
different $\epsilon$. A summary of all our gap measurements at $\nu
= 5/3$ from three samples is shown in Fig. 4(b). In our data, the
Arrhenius plots deviate from a strictly activated behavior at low
and high temperatures. This is not uncommon \cite{boebingerPRB87,
willettPRB88, shayeganPRL90} and previous studies have concluded
that the value of the energy gap can be deduced reasonably
accurately from the maximum slope in such cases.

\begin{figure}
\centering
\includegraphics[scale=1]{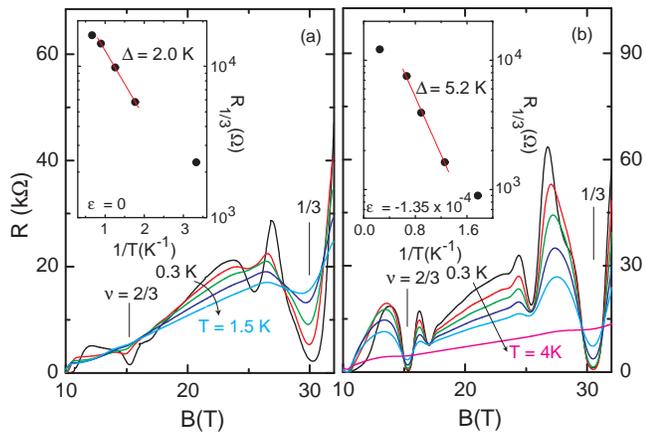}
\caption{Temperature dependence of magnetoresistance around $\nu =
1/2$ in sample A for $n = 2.4 \times 10^{11}$ cm$^{-2}$. (a) and (b) correspond to extreme values of
$\epsilon =$ 0 and $-1.35 \times 10^{-4}$. Corresponding Arrhenius plots are shown as insets.}
\end{figure}

We now discuss various aspects of our gap data, starting with $\nu =
1$ in Fig.\ 2(a). A simple fan diagram of energy vs. $\epsilon$ for
(spinless) non-interacting electrons is shown in Fig. 1(b). At a
fixed $B$, the cyclotron energy ($E_c$) is constant while the energy
levels for the [010] and [100] valleys split in response to
$\epsilon$, with the splitting given by $E_{v} = \epsilon E_{2}$. It
is clear that at $\epsilon = 0$ the state at $\nu = 1$ is expected
to be gapless. As the system is tuned away from $\epsilon = 0$, a QH
state should appear and its gap should increase linearly with a
slope of $E_{2}$ until a LL crossing occurs. After this point, the
gap is expected to be a constant equal to $E_c = \hbar e B/m_{b}$,
where $m_{b} = 0.46 m_{0}$ is the band mass for AlAs.

A similar fan diagram can be used in the case of single-valley,
two-spin systems with the role of $E_{v}$ taken over by $E_{Z}$.
Since this simple picture is inadequate to explain experimental
results, a comprehensive theory including the role of interaction
was developed for single-valley, two-spin systems
\cite{sondhiPRB93,cooperPRB97}. Here we $assume$ that the valley
degree of freedom is an isospin, and compare our experimental
results to the theory developed for the two-spin case. We point out,
however, that in our system the Fermi contours are anisotropic at
$B$ = 0. This introduces an anisotropic Coulomb interaction, the
consequences of which are unknown.

The theoretically relevant parameter in the two-spin case is the
ratio of the Zeeman energy to Coulomb energy, $\eta = E_{Z}/V_{C}$.
In our system, we redefine $\eta = E_{v}/V_{C}$. In the limit of
$\eta = 0$, the ground state at $\nu = 1$ is expected to be a QH
ferromagnet with an energy gap (for an ideal 2DES) equal to
$0.62V_{C}$ \cite{sondhiPRB93,cooperPRB97}. Experimentally
determined gaps, including the value measured in our experiments
($0.034V_{C}$ in Fig. 2(a)), however, are  much smaller
\cite{maudePRL96,shkolnikovPRL05}. Finite layer thickness, LL
mixing, and disorder are possible causes for this reduction
\cite{alaverdianPRB99,sinovaPRB00,shuklaPRB00}. We note that LL
mixing is particularly severe for electrons in AlAs because of their
large band mass. The proximity of the second LL can be appreciated
by noting that $E_{c}$ in our AlAs 2DES ($0.17 V_{C}$) is a very
small fraction of $V_{C}$.

\begin{figure}
\centering
\includegraphics[scale=1]{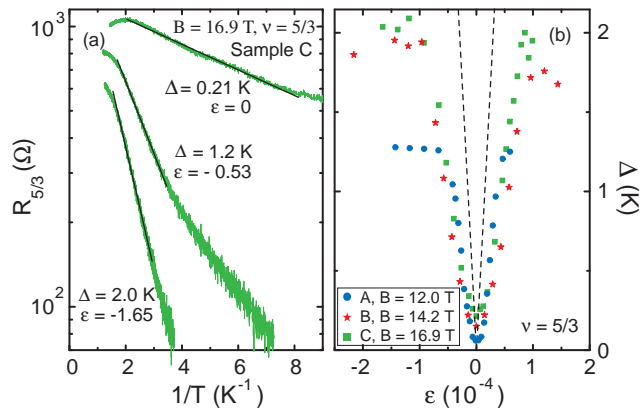}
\caption{(a) Arrhenius plots of $\nu = 5/3$ resistance minimum for
three values of $\epsilon$ (shown in units of $10^{-4}$) for sample
C at $n = 6.8 \times 10^{11}$ cm$^{-2}$. (b) Summary of $\nu = 5/3$
gap data for three samples at different $B$.}
\end{figure}

For small enough $\eta$, for a single-valley, two-spin system, the
excitations above the ferromagnetic ground state are macroscopic
spin textures (skyrmions) \cite{sondhiPRB93}. The rate at which the
$\nu = 1$ QH gap rises as a function of $E_{Z}$ gives direct
information about the size of the skyrmions
\cite{maudePRL96,schmellerPRL95,shkolnikovPRL05,shuklaPRB00}. In an
ideal 2DES, the initial slope of $\Delta$ (measured in units of
$V_{C}$) vs. $\eta$ should be infinite, then gradually decrease with
increasing $\eta$ (signaling finite-size skyrmions) and, beyond a
critical $\eta$ ($\simeq 0.054$), should be equal to one
\cite{cooperPRB97}. In Fig. 2(a) we find an initial slope of 13,
implying that if the excitations are skyrmions \cite{footnote2},
they are large but finite-sized. Finite skyrmion size has also been
reported for spin skyrmions near zero $E_{Z}$ \cite{maudePRL96}, and
is presumably because of disorder. We would like to emphasize that
the saturation value of the energy gap in Fig.\ 2(a)
($\simeq0.19V_{C}$), is close to the cyclotron energy ($0.17V_{C}$)
in our 2DES. This is not surprising since in our system we expect
the gap to be limited by the cyclotron gap (see fan diagram in Fig.
1(b)).

We now discuss our gap data at $\nu = 1/3$ (Fig.\ 2(b)).
Theoretically, for an ideal single-valley, two-spin 2DES, a
ferromagnetic ground state \cite{rezayiPRB91} with an energy gap of
about $\simeq 0.024V_{C}$ \cite{sondhiPRB93} and skyrmionic
excitations has been predicted at $\eta = 0$
\cite{sondhiPRB93,kamillaSSC96,wojsPRB02}. As in the case of $\nu =
1$, the gap is expected to quickly increase with $\eta$, signaling
the presence of skyrmions. Our measured $\nu = 1/3$ gaps, plotted in
Fig. 2(b), are qualitatively consistent with the above expectations.
The gap at $\eta = 0$ ($\simeq 0.006 V_{C}$), however, is only about
one-forth of the gap predicted for an ideal single-valley, two-spin
2DES. Also, the slope of $\Delta/V_{C}$ vs. $\eta$ is only slightly
greater than one ($\sim 1.5$) suggesting that, if the excitations
are skyrmions, their size is quite small. Finally, for very large
strains the $\nu = 1/3$ gap should saturate at a constant, expected
to be $\simeq 0.1V_{C}$ for an ideal, single-valley, single-spin
2DES \cite{halperinPRB93}. Our data of Fig. 2(b) are again
qualitatively consistent with this expectation except that the
saturation value we measure ($\simeq 0.015 V_{C}$) is much smaller
than 0.1$V_{C}$. It is likely that the much reduced values of the
gaps and the slope originate from finite-layer thickness, LL mixing,
disorder \cite{willettPRB88} and Fermi contour anisotropy. We also
remark that in our system an overall strain inhomogeneity cannot be
ruled out. At $\eta = 0$, this might lead to the formation of
ferromagnetic domains of opposite polarity. Charged excitations with
energy gaps less than that of the 2D bulk skyrmions have been
predicted at these domain walls \cite{falkoPRL99}.

Our results find a natural interpretation in light of the composite
fermion (CF) picture
\cite{jainPRL89,halperinPRB93,kamillaSSC96,CFbook} where the
fractional QHE of electrons is interpreted as the integer QHE of
CFs. Every electronic fractional filling factor $\nu$ has an integer
CF counterpart $p$. The fan diagram in Fig. 1(b) can in fact be
applied to single-spin, two-valley CFs with the modification that
$E_{c}$ denotes the cyclotron energy for CFs. Note that $\nu = 1/3$
of electrons maps to $p = 1$ of CFs, consistent with the qualitative
similarities seen in the gap data for $\nu = 1$ and 1/3 (Fig. 2). In
a two-spin system, at $\eta = 0$, interaction between the CFs is
expected to lead to a ferromagnetic QH state and skyrmionic
excitations at $\nu = 1/3$  \cite{kamillaSSC96,CFbook}. This is
qualitatively consistent with our observations . Note that according
to the fan diagram of Fig. 1(b), the saturation gap for $p = 1$ is
given by the CF cyclotron energy ($\simeq 0.1V_{C}$ for an ideal
2DES) \cite{halperinPRB93,CFbook}.

The results of our gap measurements at $\nu = 5/3$ (Figs. 2(c) and
4(b)) are qualitatively similar to the $\nu = 1/3$ data except that
the gaps are even smaller. The slope of gap vs. $\eta$ in Fig. 2(c)
is also smaller than that in Fig.\ 2(b). Within the theoretical
framework, ideally, there is no distinction between the FQH states
at $\nu = 1/3$ and 5/3 since they are related by particle-hole
symmetry. However, in real systems this symmetry is broken by
factors such as LL mixing which is particularly significant in our
system because of the large effective mass. We would like to add
that in our AlAs 2DES, at much higher strains ($|\epsilon| \gtrsim
2.3 \times 10^{-4}$ for Fig. 1(c)), we observe a crossing of the
$electron$ valley LLs as a result of which the sample resistance
near $\nu = 5/3$ dramatically increases thereby destroying the FQH
state \cite{footnote3}.

We now compare our results to those of a previous study of the role
of $spin$ splitting on the $\nu = 1/3$ QHE in GaAs
\cite{leadleyPRL97}. Leadley $et$ $al.$ measured the $\nu = 1/3$
energy gap as they tuned the $g$-factor (and thus $\eta$) through
zero via the application of hydrostatic pressure. The
magnetoresistance trace taken at $\eta = 0$ did not show a
resistance minimum at $\nu = 1/3$ but, extrapolating the gaps
measured at finite $\eta$, they concluded a finite gap ($\simeq 0.01
V_{C}$) at $\eta = 0$. In contrast, our data shows a deep resistance
minimum with a strong $T$-dependence at $\eta = 0$. Also, in Ref.
\cite{leadleyPRL97}, an initial slope of 3 was deduced from the plot
of gap vs. $\eta$, about a factor of two larger than ours. In other
relevant studies \cite{nicholasSST96,duPRB97} enhanced slopes were
reported for the $\nu = 5/3$ case, but the gaps extrapolate to a
vanishing value in the limit of $\eta = 0$.

Besides AlAs, another system in which the electrons have a valley
degree of freedom is a Si/SiGe 2DES where the two $out$-$of$-$plane$
valleys are almost degenerate. Here however, the valley splitting
cannot be easily tuned and a small valley splitting can be present
\cite{laiPRL04}. In Ref. \cite{laiPRL04} it is shown that the QH
state at $\nu = 5/3$ is absent, while the state at $\nu = 1/3$ is
present. We point out that in our AlAs samples, near $\epsilon = 0$,
we see a distinct resistance minimum at $\nu = 5/3$ only in highest
quality samples and at very low temperatures. Notably, in an 11
nm-wide AlAs quantum well \cite{bishopPRL07} at 50 mK, the
resistance minimum at $\nu = 5/3$ is almost absent when the valleys
are degenerate, suggesting a very small gap. We attribute this to
the lower sample quality where the 2DES suffers more from interface
roughness scattering due to the smaller well-width.

We thank the NSF for financial support, and J.K. Jain, S.L. Sondhi, S.A.
Parameswaran, and D. Abanin for illuminating discussions. We also
thank N.C. Bishop, G. Jones, S. A. Maier, T. Murphy, E. Palm, J.H.
Park and Y.P. Shkolnikov for assistance with the experiment. A
portion of this work was performed at the National High Magnetic
Field Laboratory, which is supported by NSF Cooperative Agreement
No. DMR-0654118, by the State of Florida, and by the DOE.

\end{document}